\documentclass[reprint,superscriptaddress,aps,prd,floatfix,nofootinbib,bibnotes]{revtex4-2}
\usepackage[colorlinks=true,linkcolor=black,citecolor=blue, urlcolor=blue,bookmarks=false]{hyperref}
\hypersetup{breaklinks=true}
\usepackage[utf8]{inputenc}
\usepackage{natbib,slashed}
\usepackage{bm}
\usepackage{dcolumn}
\usepackage{array}
\usepackage{amsmath,mathtools}
\usepackage{amssymb}
\usepackage{multirow}
\usepackage{graphicx,subcaption}
\usepackage{tensor}
\usepackage{comment}
\usepackage{enumitem}
\graphicspath{{./fig/}}
\usepackage{float}
\usepackage[normalem]{ulem}  
\usepackage{color} 

\renewcommand{\sout}{\bgroup \color{red} \ULdepth=-.5ex \ULset}

\begin{document}
\title{Diquarks and the production of charmed baryons }

\author{Hyeongock Yun}
\email{mero0819@yonsei.ac.kr}
\affiliation{Department of Physics and Institute of Physics and Applied Physics, Yonsei University, Seoul 03722, Korea}

\author{Sungsik Noh}
\email{sungsiknoh@yonsei.ac.kr}
\affiliation{Department of Physics and Institute of Physics and Applied Physics, Yonsei University, Seoul 03722, Korea}

\author{Sanghoon Lim}%
\email{shlim@pusan.ac.kr}
\affiliation{Department of Physics, Pusan National University, Pusan, Korea  }

\author{Taesoo Song}
\email{t.song@gsi.de}
\affiliation{GSI Helmholtzzentrum f\"{u}r Schwerionenforschung GmbH, Planckstrasse 1, 64291 Darmstadt, Germany}

\author{Juhee Hong}
\email{juhehong@gmail.com}
\affiliation{Department of Physics and Institute of Physics and Applied Physics, Yonsei University, Seoul 03722, Korea}

\author{Aaron Park}
\email{aaron.park@yonsei.ac.kr}
\affiliation{Department of Physics and Institute of Physics and Applied Physics, Yonsei University, Seoul 03722, Korea}

\author{Su Houng Lee}%
\email{suhoung@yonsei.ac.kr}
\affiliation{Department of Physics and Institute of Physics and Applied Physics, Yonsei University, Seoul 03722, Korea}

\author{Benjamin D\"onigus}
\email{benjamin.doenigus@cern.ch}
\affiliation{Institut f\"ur Kernphysik, Johann Wolfgang Goethe-Universit\"at Frankfurt,
Max-von-Laue-Str. 1, 60438 Frankfurt, Germany}

\date{\today}
\begin{abstract}
Utilizing a quark model characterized by parameters that effectively replicate the masses of ground state hadrons, we illustrate that $(us)$ or $(ds)$ diquarks exhibit greater compactness in comparison to $(ud)$ diquarks. Concretely, the binding energy of the $(us)$ diquark - defined as the diquark's mass minus the combined masses of its individual quarks - is found to be stronger than that of the $(ud)$ diquark.
This heightened attraction present in $(us)$ diquarks could lead to enhanced production of $\Xi_c/D$ particles in high-energy pp or ultrarelativistic heavy-ion collisions.
\end{abstract}

\maketitle

\section{Introduction}

While the main objective of the heavy-ion collision experiments at CERN is to study the properties of quark-gluon plasma~\cite{Andronic:2017pug}, recent results including those from pp and pPb collisions have provided new opportunities to study QCD and hadron properties in general~\cite{ALICE:2022wpn}.  Of particular interest are the recent measurements of heavy baryon to heavy meson ratios in pp, pPb and PbPb collisions because these probe the quark-quark interaction in the hadronization processes~\cite{ALICE:2021bib,ALICE:2021psx,ALICE:2021bli}.

The baryon-to-meson ratio enhancement for both the light and heavy quark sectors observed in heavy-ion collision can be explained by medium effects in the hadronization process well encoded in the coalescence model~\cite{Greco:2003xt,Plumari:2017ntm,Cho:2019lxb,Beraudo:2022dpz}. What is interesting to see is that  $\Lambda_c/D$  enhancement over that expected from the fragmentation process exists even in pp collision, which can only be explained when one assumes an additional production mechanism for the baryon\cite{Beraudo:2023nlq}. In the thermal model additional resonances not listed by the Particle Data Group~\cite{ParticleDataGroup:2022pth} are needed~\cite{Andronic:2021erx,ALICE:2022wpn}. Even more striking, the models that explain the $\Lambda_c/D$ enhancement underestimate the recent measurement of the $\Xi_c/D$ ratio\cite{Minissale:2023xdj}.  

In this work, we will show that a strong $(qs)$ diquark correlation, where $q=u,d$, provides a new effect that provides an additional production mechanism to enhance the $\Xi_c/D$ ratio.   

The two light quarks in the ground state baryon with one heavy quark are in the color anti-triplet channel with either isospin zero (spin zero) or isospin one (spin one), which respectively form the lowest triplet or sextet charmed baryon representations in flavor SU(3), respectively.   Hence, if strong diquark correlations exist, a strong overlap of the diquarks and a heavy quark could lead to additional enhancement in the heavy baryon-to-meson ratio.  On the other hand, data seems to suggest that there should be an additional enhancement in the $\Xi_c/D$ ratio compared to $\Lambda_c/D$ ratio in pp collision at both $\sqrt{s}=5.02$ TeV and 13 TeV~\cite{ALICE:2021psx,ALICE:2021bli}.  The diquark in the $\Xi_c$ and $\Lambda_c$ are $(qs)$ and $(ud)$ diquarks, respectively.  As we will see, the quark model that well reproduces the ground state hadron masses shows that there is a stronger attraction in the $(qs)$ diquark compared to that in the $(ud)$ diquark.  This will lead to a small but non-negligible enhancement in $\Xi_c$ productions.  Enhanced production in the presence of attraction can be shown to be true in general using S-matrix theory for hadron production~\cite{Huovinen:2016xxq}. Here, we will employ a coalescence model to estimate the extent of the additional enhancement anticipated in the $\Xi_c/D$ ratio as a result of the strong $(qs)$-diquark correlation, which is represented by a corresponding diquark distribution inherent in the hadronization process

The paper is organized as follows.  In Sec. II, we discuss the strength and the binding energies of the $(us)$ and $(ud)$ diquarks.  We then construct a phenomenological coalescence model to estimate the additional $\Xi_c/D$ ratio production in pp, pPb, and PbPb collisions. Finally, we give the summary.  

\section{Diquarks in quark model}

It has been long noticed that diquarks play important roles in hadron structures and reactions~\cite{Anselmino:1992vg}.  
To study the flavor dependence on their binding, let us study the diquark configurations in a quark model.  Specifically, we evaluate the masses of diquarks by using a non-relativistic quark model described by the following Hamiltonian\cite{Park:2018wjk}.
\begin{eqnarray}
H &=& \sum^{n}_{i=1} \left( m_i+\frac{{\mathbf p}^{2}_i}{2 m_i} \right)-\frac{3}{4}\sum^{n}_{i<j}\frac{\lambda^{c}_{i}}{2} \,\, \frac{\lambda^{c}_{j}}{2} \left( V^{C}_{ij} + V^{CS}_{ij} \right), \qquad
\label{Hamiltonian}
\end{eqnarray}
where $n=2(3)$ for diquarks (baryons). The internal quark potentials $V^C_{ij}$ and $V^{CS}_{ij}$ in Eq.~(\ref{Hamiltonian}) are as follows.
\begin{eqnarray}
V^{C}_{ij} &=& - \frac{\kappa}{r_{ij}} + \frac{r_{ij}}{a^2_0} - D,
\label{ConfineP}
\\
V^{CS}_{ij} &=& \frac{\hbar^2 c^2 \kappa'}{m_i m_j c^4} \frac{e^{- \left( r_{ij} \right)^2 / \left( r_{0ij} \right)^2}}{(r_{0ij}) r_{ij}} \boldsymbol{\sigma}_i \cdot \boldsymbol{\sigma}_j\,,
\label{CSP}
\end{eqnarray}
where
\begin{eqnarray}
r_{0 ij} &=& 1/ \left( \alpha + \beta \frac{m_i m_j}{m_i + m_j} \right)\,,
\nonumber \\
\kappa' &=& \kappa_0 \left( 1 + \gamma \frac{m_i m_j}{m_i + m_j} \right)\,.
\nonumber
\end{eqnarray}
This model can be used to fit the ground state hadron masses including light, charm, and bottom quarks\cite{Park:2018wjk,Noh:2021lqs}.  The model has also been extensively used to study possible compact exotic configurations\cite{Park:2018wjk,Noh:2021lqs,Park:2017jbn}.
Here, we fit the model parameters in the Hamiltonian to the ground state masses listed in Table~\ref{baryons}.  The list is limited to baryon states to better fit the quark-quark interactions with different flavors. 
The selected baryons in Table~\ref{baryons} contain the diquark structure of interest in this work. The fitted model parameters are as follows.
\begin{eqnarray}
&\kappa=130.0 \, \textrm{MeV fm}, \quad a_0=0.0318119 \, \textrm{(MeV$^{-1}$fm)$^{1/2}$}, &
\nonumber \\
&D=975  \, \textrm{MeV}, \quad m_{u}=m_{d}=335 \, \textrm{MeV}, & \nonumber \\
&m_{s}=652 \, \textrm{MeV}, \quad
m_{c}=1940 \, \textrm{MeV}, 	&\nonumber \\
&\alpha = 1.2499 \, \textrm{fm$^{-1}$}, \,\, \beta = 0.0008314 \, \textrm{(MeV fm)$^{-1}$}, &	\nonumber \\
&\gamma = 0.00168 \, \textrm{MeV$^{-1}$}, \,\, \kappa_0=185.144 \, \textrm{MeV}. 	 &
\label{parameters1}
\end{eqnarray}
\begin{table}[t]
\caption{Masses of baryons obtained (Column 3) from the model calculation in this work with the fitting parameter set given in  Eq.~(\ref{parameters1}).  Column 4 shows the variational parameters $a_1$ and $a_2$.}

\centering

\begin{tabular}{cccc}
\hline
\hline	\multirow{2}{*}{Particle}	&	Experimental	&	Mass		&	\quad Variational\quad					\\
								&	Value (MeV)	&	(MeV)	&	\quad Parameters (${\rm fm}^{-2}$)\quad	\\
\hline  
$\Lambda$		&	1115.7	&	1116.3	&	\quad$a_1$ = 3.0, $a_2$ = 2.9\quad	\\
$\Lambda_{c}$	&	2286.5	&	2272.2	&	\quad$a_1$ = 3.1, $a_2$ = 3.9\quad	\\
$\Sigma_{c}$		&	2452.9	&	2446.2	&	\quad$a_1$ = 2.2, $a_2$ = 4.0\quad	\\
$\Sigma_{c}^*$	&	2517.5	&	2531.4	&	\quad$a_1$ = 2.0, $a_2$ = 3.5\quad	\\
$\Sigma$			&	1192.6	&	1202.2	&	\quad$a_1$ = 2.2, $a_2$ = 3.3\quad	\\
$\Sigma^*$		&	1383.7	&	1401.1	&	\quad$a_1$ = 1.9, $a_2$ = 2.5\quad	\\
$\Xi$			&	1314.9	&	1331.8	&	\quad$a_1$ = 3.6, $a_2$ = 3.1\quad	\\
$\Xi^*$			&	1531.8	&	1544.1	&	\quad$a_1$ = 3.1, $a_2$ = 2.3\quad	\\

$\Xi_{c}$		&	2467.8	&	2474.2	&	\quad$a_1$ = 3.5, $a_2$ = 5.0\quad	\\
$\Xi_{c}^*$		&	2645.9	&	2654.9	&	\quad$a_1$ = 2.6, $a_2$ = 4.6\quad	\\
$\Xi'_{c}$		&	2579.2	&	2570.2	&	\quad$a_1$ = 2.8, $a_2$ = 5.2\quad	\\
$\Omega_{c}$		&	2695.2	&	2684.7	&	\quad$a_1$ = 3.9, $a_2$ = 6.1\quad	\\
$\Omega_{c}^*$	&	2765.9	&	2768.8	&	\quad$a_1$ = 3.6, $a_2$ = 5.3\quad	\\

$p$				&	938.27	&	951.42	&	\quad$a_1$ = 2.5, $a_2$ = 2.5\quad	\\
$\Delta$			&	1232	&	1246.9	&	\quad$a_1$ = 1.8, $a_2$ = 1.8\quad	\\
\hline 
\hline
\label{baryons}
\end{tabular}
\end{table}
The two variational parameters  $a_1$ and $a_2$ appearing in Table \ref{baryons} are the scaling factors of the Gaussian wave function for the relative distance between the two light quarks within the diquark and the relative distance between the center of the diquark and the heavy quark, respectively.
The standard deviation of the masses obtained in Table~\ref{baryons} is $
\sigma = ( \frac{1}{N-1} \sum_{i=1}^{N} \left( M^{Thr}_i - M^{Exp}_i \right)^2 )^{1/2} = 11.9~{\rm MeV},$ where  $M^{Thr}_i$ indicates the mass obtained by the model calculation and $M^{Exp}_i$ the experimentally measured mass.

\begin{table}[t]

\caption{Mass, binding energy, and size of $ud$, $us$, and $uc$ diquarks in a color $\mathbf{\bar{3}}$ and spin 0 state. The binding energy is defined as $M_{Diquark}-m_{u} - m_{q}$ where $m_{q}$ can be $m_{d}$, $m_{s}$, or $m_{c}$.}

\centering

\begin{tabular}{ccccccc}
\hline
\hline
Diquark	&&	Mass(MeV)&&Binding energy(MeV)&&	Size(fm)\\
\hline 
$ud$ 	&&	680.3	&&	10.22	&&	0.761	\\
$us$ 	&&	970.7	&&	-16.30	&&	0.714	\\
$uc$ 	&&	2244.8	&&	-30.23	&&	0.700	\\
\hline 
\hline
\end{tabular}
\label{Diquarks1}
\end{table}

\begin{table}[t]

\caption{Individual contribution to the mass of $ud$, $us$, and $uc$ diquarks in Table~\ref{Diquarks1}. In the following table, $m_{q}$ can be $m_{d}$, $m_{s}$, or $m_{c}$. All units are MeV.}

\centering

\begin{tabular}{cccccccc}
\hline
\hline
	Contribution					&&	$ud$		&&	$us$		&&	$uc$		\\
\hline 
$m_{u} + m_{q} - \frac{1}{2}D$	&&	182.5	&&	499.5	&&	1787.5	\\
$\sum^{2}_{i=1} \frac{{\mathbf p}^{2}_i}{2 m_i}$
					 			&&	383.6		&&	329.9	&&	265.8	\\
				$V^C$ 			&&	267.1		&&	236.6	&&	227.5	\\
				$V^{CS}$		 	&&	-152.9		&&	-95.34	&&	-36.0	\\
\hline
Total	&&	680.3	&&	970.7	&&	2244.8		\\
\hline 
\hline
\end{tabular}
\label{Diquarks2}
\end{table}

We now calculate the masses of the $(ud)$ and $(us)$ diquarks $M_{Diquark}$ using these parameters.  The results are summarized in Table~\ref{Diquarks1}.
In the table, one can see that the binding is stronger for the heavier diquarks. 
The binding energy is defined as $M_{Diquark}-m_{u} - m_{i}$ where $m_{i}$ can be $m_{d}$, $m_{s}$, or $m_{c}$.
Also, one can see that the size of the diquark becomes smaller when the component quark becomes heavier. To analyze the origin of the stronger binding for the heavier diquarks, 
we show each part of the Hamiltonian contributing to the masses of diquarks in Table~\ref{Diquarks2}. 
We first note that the attraction coming from the color-spin interaction $V^{CS}$ becomes smaller as the quark masses increase.  This is due to the inverse quark mass dependence in 
$V^{CS}$. On the other hand, both the Coulomb and confining potential $V^C$ decrease as the quark masses increase.  
This is attributed to the reduced size of heavier diquarks. Heavier quarks experience a more pronounced Coulomb attraction and a weaker linearly rising potential, a phenomenon commonly observed when dealing with heavier quarks, as discussed previously in Ref. \cite{Karliner:2014gca,Karliner:2017qjm}.
Altogether, one finds that the diquarks become more bound as the quarks involved become heavier.
As a result, in Table~\ref{Diquarks1}, $(us)$ and $(uc)$ diquarks are bound while $(ud)$ diquark is not.

\subsection{Diquark mass at finite temperature}

In reality, one has to introduce a model scenario to implement the strong $(us)$ correlation at the hadronization point in the presence of quarks. 
One potential scenario involves assuming the presence of local quark matter at the point of hadronization, investigating the persistence of diquark correlations, and subsequently assessing the impact of diquarks on the process of hadronization.

For that purpose, we will first analyze the diquarks when the Coulomb and the confining part of the potential are modified as given in Ref.\cite{Lafferty:2019jpr} at the chemical freeze-out temperature.  Then the color potential in Eq. \eqref{ConfineP} will be modified as follows.
\begin{eqnarray}
V^{C}_{ij} &=& - \kappa \bigg[m_D+\frac{e^{-m_D r_{ij}}}{r_{ij}} \bigg] \nonumber \\
&&+ \frac{1}{a^2_0} \bigg[
\frac{2}{m_D}-\frac{e^{-m_Dr_{ij}}(2+m_D r_{ij})}{m_D}
\bigg] - D,
\label{ConfineP-T}
\end{eqnarray}
which reduces to Eq. \eqref{ConfineP} when the Debye screening mass $m_D\rightarrow 0$.  Here we will consider $m_D$ given in Ref. \cite{Lafferty:2019jpr} at several temperatures above the critical point.

\begin{table}[h!]

\caption{Differences in binding energies $B_{(us)} - B_{(ud)}$ where $B_{(us)}$ and $B_{(ud)}$ indicate the binding energies of $(us)$ and $(ud)$ diquarks, respectively. The mass and size of $(us)$ diquark are also presented. The table presents the results obtained from five different schemes discussed in the text.}

\centering

\begin{tabular}{c|cccccc}
\hline
\hline
Type	&	$B_{(us)} - B_{(ud)}$	&	$m_{us}$(MeV)  &	Size(fm)\\
\hline 
Scheme 1	&	-26.52 	&	970.7	&	0.714	\\
\hline
Scheme 2	&	-11.14 	&	710.5	&	0.753	\\
\hline 
Scheme 3	&	-9.67 	&	673.2	&	0.800	\\
\hline 
Scheme 4	&	-7.81 	&	613.8	&	0.889	\\
\hline 
Scheme 5	&	-6.93 	&	580.2	&	0.954	\\
\hline 
\hline
\end{tabular}
\label{Diff_udus}
\end{table}

For the light-quark system, we will further consider the effects of chiral symmetry restoration and thermal masses. To this end, in evaluating the masses of diquarks within our quark model approach, we will study cases where we use the thermal masses for the light quarks, typically taken to be 300 MeV for the $u,d$ quarks, and 400 MeV for the strange quark, which is about 100 MeV heavier than the light quark mass as is the case for the bare quark masses. 

We now solve for the diquark masses and their bindings for the following five cases.
1) Scheme 1(Table~\ref{Diquarks1}): use the potential and quark masses as fitted to the baryon mass spectrum. That is, use zero temperature
potential in Eq.~(\ref{ConfineP}) with $m_s = 652$ MeV and $m_u = 335$ MeV.
2) Scheme 2: use the zero temperature potential in Eq.~(\ref{ConfineP}) but with $m_s = 400$ MeV and $m_u = 300$ MeV.
3) Scheme 3: use the thermal potential in Eq.~(\ref{ConfineP-T}) with $m_s = 400$ MeV, $m_u = 300$ MeV, and $m_D = T_c =156$ MeV.
4) Scheme 4: use the thermal potential in Eq.~(\ref{ConfineP-T}) with $m_s = 400$ MeV, $m_u = 300$ MeV, and $m_D \sim 1.5\times T_H$ MeV\cite{Lafferty:2019jpr}, where we take $T_H=181$ MeV to be the hadronization temperature in pp collision from the flow analysis given later.

We also introduce Scheme 5, aimed at describing the phase where chiral symmetry is restored. 
From a phenomenological point of view, the value of the potential at $r\rightarrow \infty$ is related to the creation of the quark-antiquark pair, whose mass is related to chiral symmetry breaking\cite{Gubler:2020hft}. 
When $m_D$ is small, the potential is very large due to the linearly rising potential. That is the point where the usual quark model is applied to calculate the ground state hadron masses; the potential at large values does not affect the properties of the ground state hadrons as they are smaller than 1 fm in size.  On the other hand, $m_D$ increases with temperature and when $m_D \sim 0.325 $ GeV,  $ V(r \rightarrow \infty) =0$.  Lattice gauge theory fit to $m_D$  indeed indicates that 0.325 GeV is reached near the critical temperature within the lattice error\cite{Lafferty:2019jpr}. 
Therefore, we consider Scheme 5, where we take $m_D=325$ MeV and take the quark masses to be their thermal masses
$m_s = 400$ MeV and $m_u = 300$ MeV.

Table~\ref{Diff_udus} shows the result for the differences in the binding energy between the $(us)$ and $(ud)$ diquarks, the mass of the $(us)$ diquark, and the size of the $(us)$ diquark for the five schemes.  
As can be seen in the differences in the bindings, for all cases, the $(us)$ diquark has a stronger attraction than the $(ud)$ diquark.  We further find that within the uncertainties given in the present analysis, the $(us)$ diquark mass could be between 580 to 970 MeV. Subsequently, in the following section, we will assess the influence of the $(us)$ diquark using the coalescence model and incorporating the estimated range of $(us)$ diquark masses.

\subsection{Diquarks and hadron overlap}

The strong correlation within the $(us)$ diquark leads to increased production when a light quark $q$ and an $s$-quark are in close proximity; this correlation is stronger.  These effects should be incorporated into the modeling of hadron production. As previously discussed, the light diquark within the baryon triplets $\Lambda_c$ and $\Xi_c$ is primarily the so-called 'good diquark,' characterized by an anti-triplet color configuration, zero spin, and anti-triplet flavor.  
An explicit quark model calculation reveals that the SU(3) flavor symmetry breaking in $\Xi_c$ results in a mixing of less than 0.01\% from the spin-1 diquark component.   
This implies that we can indeed assume the $(us)$ diquark within the $\Xi_c$ to be a good diquark, denoted as $(us)_{S=0}$ with the subscript indicating a spin of 0. 
Assuming multiple quarks are present in the hadronization process, the strong correlation between a light quark $q$ and a strange quark $s$ characterized by $(qs)_{S=0}$ will 
provide an additional production mechanism for $\Xi_c$.  

It should be noted that the strong correlation of $(qs)_{S=0}$ has a smaller effect on the production of baryon octets with strange quarks as the overlaps are smaller.  
This is because the spin-1/2 baryon octet is composed of a mixed flavor symmetry so that any two quarks within a baryon will be in either a color anti-triplet and flavor symmetric configuration or a color anti-triplet and flavor antisymmetric configuration with equal probability. In other words, the quark model predicts that the probability of a diquark being in the "good" diquark state inside any baryon octet state is 1/2, while inside a heavy baryon, it is 1.
Therefore, focusing on the $(qs)$ diquark, we find the following constraint.

\begin{eqnarray}
\sum_{q=u,d}| \langle (sq)_{{S=0}}|B_8 \rangle|^2 & \leq & \frac{1}{2},   \\
\sum_{q=u,d}|\langle (sq)_{{S=0}}|\Xi_c \rangle|^2 & =& 1, 
\end{eqnarray}
where  $| B_8 \rangle $   ($ | \Xi_c \rangle$ )  means any diquark in the flavor octet baryon ($\Xi_c$) state.  
Table \ref{Diquarks3} shows how all the probabilities of diquarks inside the baryon octet states are distributed.
For example,  $|    \langle (us)_{{S=0}}|\Sigma^+ \rangle|^2  = \frac{1}{2}$ and $|    \langle (us)_{{S=0}}|\Sigma^0 \rangle|^2 =|    \langle (ds)_{{S=0}}|\Sigma^0 \rangle|^2= \frac{1}{4}$, while $|    \langle (us)_{{S=0}}|\Lambda \rangle|^2 = \frac{1}{6}$.  Also,  $|    \langle (qs)_{{S=0}}|\Xi \rangle|^2  = \frac{1}{2}$. 
Consequently, the presence of a $(us)_{S=0}$ will have a lesser impact on the production of hyperons compared to the anticipated effect on the production of $\Xi_c$. Thus, we will solely focus on the supplementary production of $\Xi_c$ using the coalescence model.

\begin{widetext}

\begin{table}[t]

\caption{Probabilities of each diquark component for the baryon octet states are depicted. Each diquark configuration is represented in the second row by the respective color-spin factor, considering the quark masses of the two quarks $i$ and $j$ in the diquark as $-\frac{1}{m_i m_j} \lambda_i^c \lambda_j^c \sigma_i \cdot \sigma_j$. The color-spin factors in the second column are determined by multiplying the probability of a diquark with the corresponding color-spin factors containing mass terms multiplied by 3. These values are summed across all diquarks listed in the row for the corresponding baryon}

\centering
\begin{tabular}{|c|c|ccc|ccc|}
\hline
&&  \multicolumn{3}{c|}{Good diquarks} &\multicolumn{3}{c|}{Bad diquarks} \\
\cline{3-8}
 Baryon & Color-spin factor & $-\frac{8}{m_q^2}$ &	 
 $-\frac{8}{m_qm_s} $ & $-\frac{8}{m_s^2} $ &$\frac{8}{3m_q^2}$ &	 
 $\frac{8}{3m_qm_s} $ & $\frac{8}{3m_s^2} $ \\
\hline 
$p,n$ & $-\frac{8}{m_q^2}$	& $\frac{1}{2}$ & 0	& 0 &	$\frac{1}{2}$	& 0 & 0	\\
$\Lambda$ & $-\frac{8}{m_q^2}$	& $\frac{1}{3}$ & $\frac{1}{6}$	& 0 &	0	& $\frac{1}{2}$ & 0	\\
$\Sigma$ & $\frac{8}{3m_q^2}-\frac{32}{3m_q m_s}$	& 0 & $\frac{1}{2}$	& 0 &	$\frac{1}{3}$	& $\frac{1}{6}$ & 0	\\
$\Xi$  & $\frac{8}{3m_s^2}-\frac{32}{3m_q m_s}$	&  0 & $\frac{1}{2}$	& 0 &	0	& $\frac{1}{6}$ & $\frac{1}{3}$	\\
\hline
\end{tabular}
\label{Diquarks3}
\end{table}

\end{widetext}

\subsection{Diquarks with flow}

To incorporate the strong correlation in the $(qs)$ diquark within a coalescence model, we will assume the presence of $(qs)$ diquarks in the quark-gluon plasma\cite{Lee:2007wr} and estimate their total number using the thermal model.
Whether to explicitly include all or part of the diquark configurations is a subtle question. 
In coalescence models of hadron production from the quark-gluon plasma, the 3-body coalescence formula for $\Lambda_c$ production, as presented in Refs. \cite{Plumari:2017ntm,Cho:2019lxb}, involves phase space integrals of two relative coordinates. Consequently, it can be reformulated as a quark-quark to diquark 2-body coalescence, along with an additional diquark-heavy quark 2-body coalescence. 
Hence, with suitable normalization factors, one can effectively replace 3-body coalescence with the 2-body coalescence with diquarks if there is no binding energy.  
This seems to be consistent with coalescence approaches, where the bulk part of the heavy baryon production comes from the recombination of the heavy quark and a surrounding diquark\cite{Beraudo:2022dpz,Beraudo:2023nlq}.  
What we want to 
emphasize is that there is an additional binding in the (qs) diquark, which has to be taken into account through the existence of bound diquarks,  when using the coalescence model\cite{Lee:2007wr}, to explain the missing strength in the $\Xi_c/D^0$ production.  

The expected diquark number at central rapidity can be estimated using the statistical model at the hadronization temperature $T_H $~\cite{Vovchenko:2019kes}.
\begin{align}
N^{stat}_{[us]} &= V_H \frac{g_{[us]}}{2\pi^2} \int^{\infty}_0 \frac{p^2 dp}{\gamma^{-1}_h e^{E_h / T_H} \pm 1} ,
\label{shm}
\end{align}
where $V_H$ is the volume at the chemical freeze-out point. Later, we will use the limiting volume $V_H=20$ fm$^3$ extracted for a small system in pp collision at 7 TeV ~\cite{Mazeliauskas:2019ifr}.  
We further augment the distribution with the flow. 
The collective radial expansion of the fireball, created in heavy-ion collisions can be understood well within a hydrodynamic picture. A more phenomenological way to capture this isotropic expansion is known as blast-wave model~\cite{Schnedermann:1993ws}. 
The model assumes a spectrum of purely thermal sources which are boosted in transverse direction. The velocity distribution in $0 \leq r \leq R$ is assumed to be 
\begin{eqnarray}
\beta_r=\left(\frac{r}{R} \right)^n \beta_s,
\label{beta}
\end{eqnarray}
where $\beta_s$ is the surface velocity, a free parameter of the fit. In many applications, a linear profile is assumed and $n$ is fixed equal to unity. The quality of the fit can be improved if $n$ is considered as an additional free parameter. The resulting values for the kinetic freeze-out temperature $T_{kin}$ and $\beta_s$ are generally anti-correlated. The so-obtained spectral shape is a superposition of the contributions due to the individual thermal sources and is given by
\begin{equation}
\frac{1}{m_{\rm T}}\frac{{\rm d}N}{{\rm d}m_{\rm T}} \propto m_{\rm T}\int_0^R I_0 \left( \frac{p_{\rm T} \sinh \rho}{T_{kin}} \right) K_1 \left( \frac{m_{\rm T} \cosh \rho}{T_{kin}} \right) r\, {\rm d}r \,,
\label{equation:blastwave}
\end{equation}
where $I_0(x)$ and $K_1(x)$ are Bessel functions, $m_{\rm T}=\sqrt{p_{\rm T} ^2 + m^2}$ and $\rho = \tanh^{-1} \beta_r$.

\begin{figure}[!htb]
\begin{center}
\includegraphics[width=0.49\textwidth]{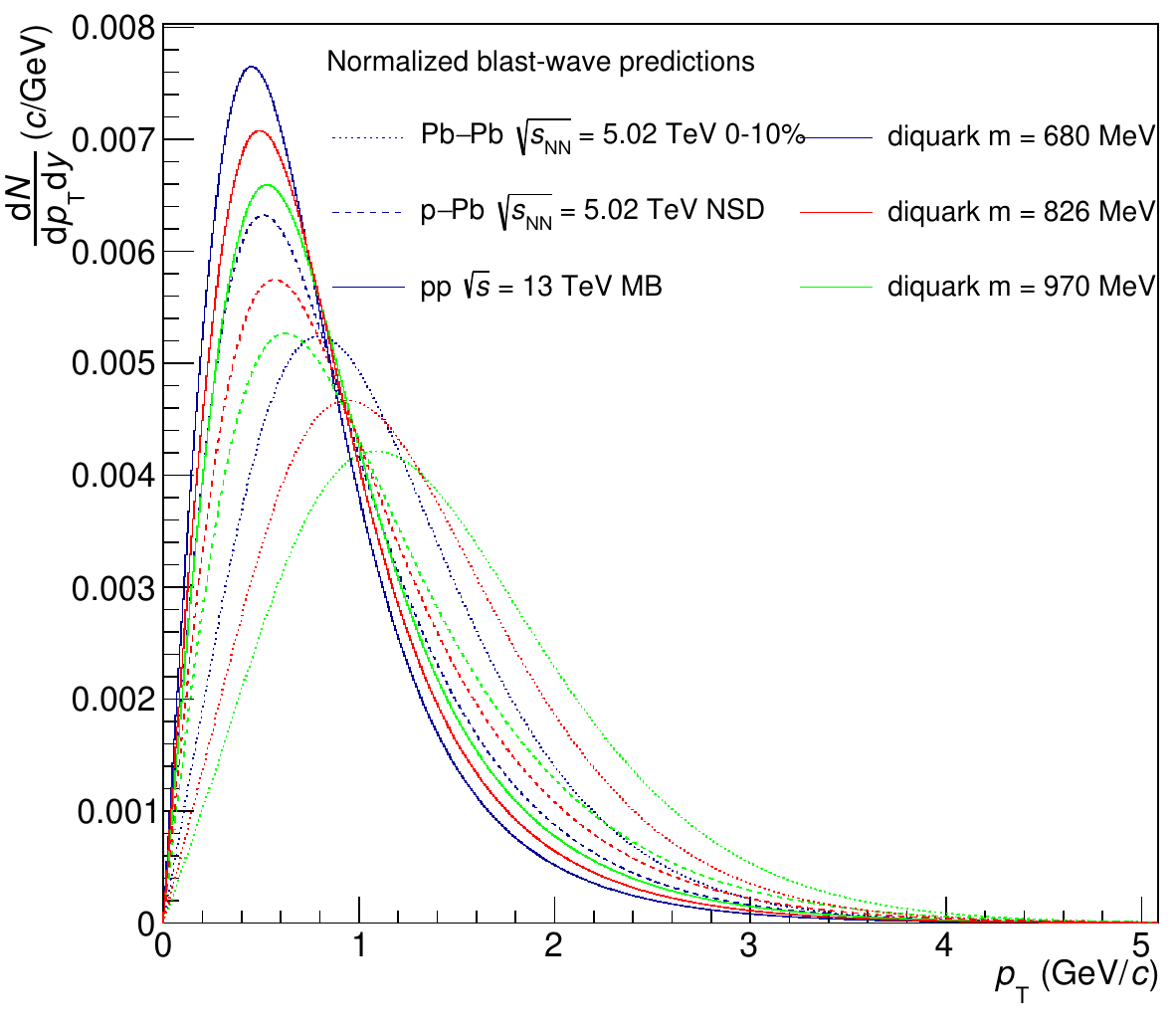}
\caption{\label{fig:bw_prediction} Normalised blast-wave distributions for three different possible diquark masses (680 MeV, 826 MeV and 970 MeV) and for three collisions systems (pp minimum bias, p--Pb non-single diffractive and central Pb--Pb). }
\end{center}
\end{figure}

Blast-wave fits to existing data have been performed by the ALICE Collaboration at the LHC for different collision energies and collision systems, spanning a large range of charged particle multiplicities per pseudorapidity $\langle {\mathrm d}N_{\mathrm ch}/{\mathrm d}\eta\rangle $ per event. A summary of this is discussed in Ref.~\cite{ALICE:2020nkc}. The typical approach is to fit the transverse-momentum spectra of charged $\pi$, K, and p altogether and by that extract a common freeze-out temperature $T_{kin}$ and mean velocity $\langle \beta \rangle$. In addition, a shape factor $n$ is used.
This set of common blast-wave parameters can be used to predict the shape of spectra of unmeasured particles (spectra). It has been used by the ALICE Collaboration to encounter possible biases in case only parts of the whole $p_{\mathrm T}$ spectrum was accessible~\cite{ALICE:2017jmf} or something similar~\cite{ALICE:2015udw}.
We use it here to predict the transverse momentum shape for calculations where flow is not included. The normalized blast-wave function can be multiplied onto the corresponding number for the constituent with the same mass to obtain the transverse momentum distribution. The parameters for the blast-wave functions for four multiplicity values, corresponding to minimum bias pp collision at $\sqrt{s}= 13$ TeV, minimum bias pp collision at $\sqrt{s}= 5$ TeV, non-single diffractive p--Pb at $\sqrt{s_{\mathrm {NN}}}= 5$ TeV and central Pb--Pb collisions at $\sqrt{s_{\mathrm {NN}}}= 5$ TeV, are given in Table \ref{tab:values}.
\begin{table}[!htb]
\centering
\begin{tabular}{|c|c|c|c|c|}
\hline
Collision system & $\langle{\mathrm d}N_{\mathrm ch}/{\mathrm d}\eta\rangle$ & $T_{kin}$ (GeV) & $\langle\beta\rangle$ & n \\
\hline
\hline
pp 13 TeV MB & 6.88 & 0.184 & 0.270 & 3.878 \\
\hline
pp 5.02 TeV MB & 4.30 & 0.181 & 0.198 & 6.248 \\
\hline
p--Pb 5.02 TeV NSD & 17.81 & 0.177 & 0.423 & 1.846 \\
\hline
Pb--Pb 5.02 TeV 0-10\% & 1781 & 0.113 & 0.659 & 0.650 \\
\hline
\end{tabular}
\caption{Values used for the blast-wave predictions shown in Fig.~\ref{fig:bw_prediction}}
\label{tab:values}
\end{table}

We can now predict the transverse momentum shape of the diquarks with different masses in the blast-wave model with parameters given in Table \ref{tab:values} for different multiplicities.  The results are shown in Fig.~\ref{fig:bw_prediction} for three different diquark masses. 

\section{$\Xi_c/D^0$ ratios}

To estimate the additional enhancement of $\Xi_c/D^0$ ratio coming from $(us)$ diquark, we use the 2-dimensional coalescence model. 

\begin{widetext}

\begin{align}
\frac{\mathrm{d}^2 N_{\Xi_c^+}}{\mathrm{d}^2 P_{\mathrm{T}}} =  \frac{g_{\Xi_c^+} }{g_{[us]}g_c} & \int d^2x_1 d^2 x_2 d^2 p_{1\mathrm{T}} d^2 p_{2\mathrm{T}} \ \frac{\mathrm{d}^2 N_{[us]}}{A_L \mathrm{d}^2 p_{[us]\mathrm{T}} } \frac{\mathrm{d}^2 N_c}{A_L \mathrm{d}^2 p_{c\mathrm{T}} } 
\times
W_{\Xi_c^+}(\vec{r}, \vec{k}) \delta^{(2)} ( \vec{P}_{\mathrm{T}} - \vec{p}_{[us]\mathrm{T}} - \vec{p}_{c\mathrm{T}} )
\nonumber \\
=\frac{ g_{\Xi_c^+} }{g_{[us]}g_c}  &\left( 2\sqrt{\pi} \sigma \right)^2 \frac{1}{A}  \int d^2 p_{1\mathrm{T}} d^2 p_{2\mathrm{T}} \ \frac{\mathrm{d}^2 N_{[us]}}{ \mathrm{d}^2 p_{[us]\mathrm{T}} } \frac{\mathrm{d}^2 N_c}{\mathrm{d}^2 p_{c\mathrm{T}} } 
 \times
 \exp{\left[ -\sigma^2 p'^2 \right]}  \delta^{(2)} ( \vec{P}_{\mathrm{T}} - \vec{p}_{[us]\mathrm{T}} - \vec{p}_{c\mathrm{T}} ),
\label{xi_coal}
\end{align}
\end{widetext}
where the $g$ values and $A$ are statistical factors and the coalescence area at freeze-out point. In this model, we used the following two-dimensional Gaussian-type Wigner function.  

\begin{align}
W_{\Xi_c^+} = 4\exp{\left( -\frac{r'^2}{\sigma^2} - \sigma^2 k'^2 \right) }.
\label{Wigner}
\end{align}
The parameter $\sigma$ is related to the $\Xi_c^+$ radius by $\sigma = \sqrt{8/3}r_{\Xi_c^+}$ and we used  $r_{\Xi_c^+}=0.222$ fm  obtained by quark model calculation.  Furthermore, the primed momenta in Eq.\eqref{xi_coal} and Eq.\eqref{Wigner} are taken in the center-of-mass frame of the $\Xi_c$, as discussed in Ref.~\cite{Cho:2019lxb}.

To determine $A$, we use the method in Ref. \cite{Yun:2022evm}.  The yield of d and $^3$He in PbPb collisions are determined at the chemical freeze-out point~\cite{Andronic:2017pug}. Therefore, we can obtain the coalescence area $A$ at the chemical freeze-out point by fitting the production data at pp collisions using the same coalescence formula given in Eq. \eqref{xi_coal} for the deuteron and the corresponding three-body formula for $^3$He. It should be noted that in the case of pp collisions, where the kinetic freeze-out temperature exceeds the critical temperature, we consider the chemical freeze-out temperature to be the same as the kinetic freeze-out temperature.

\begin{figure}[h]
\centering
\includegraphics[width=.45\linewidth]{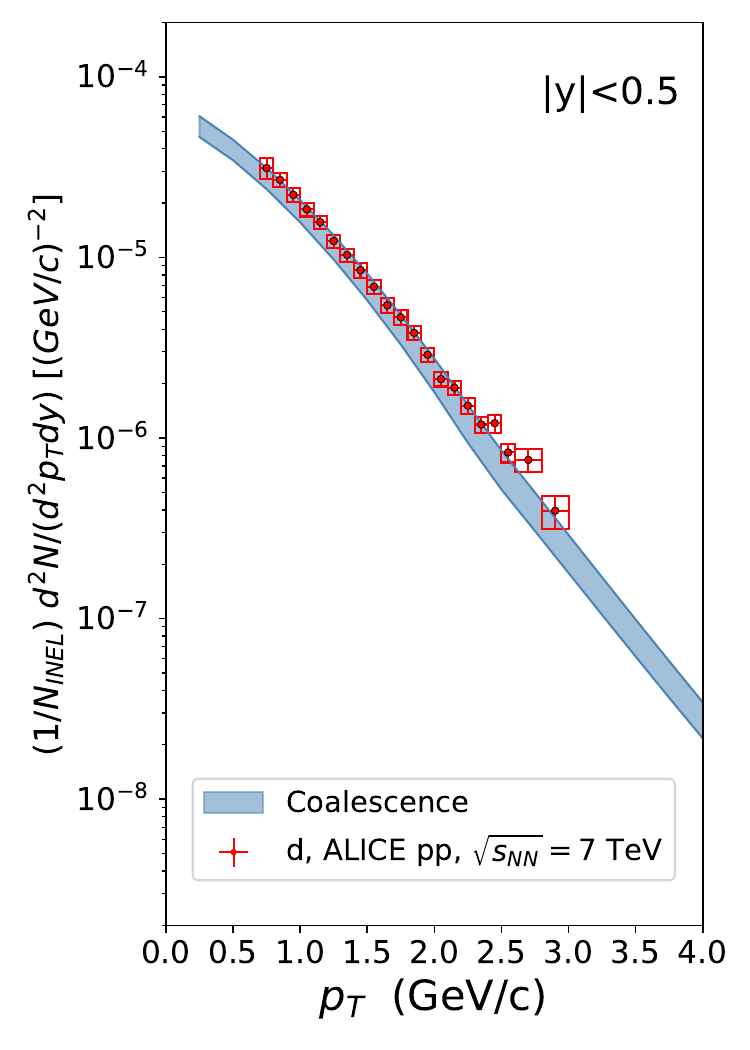}
\includegraphics[width=.45\linewidth]{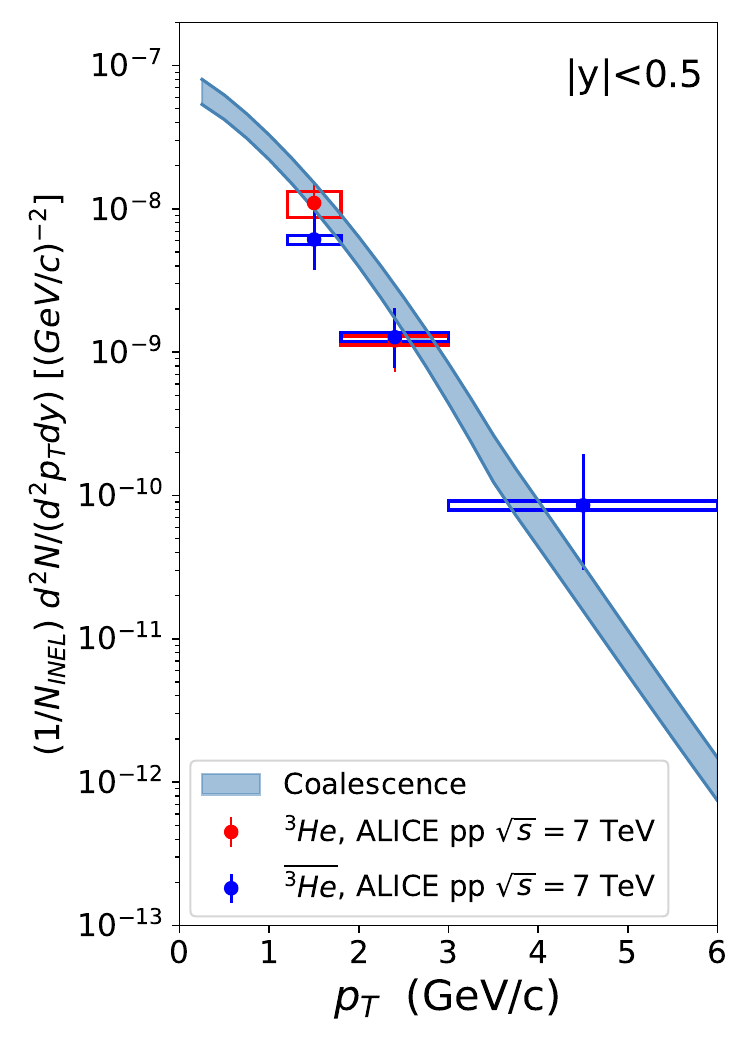}
\caption{Transverse momentum distribution of d (left panel) and $^3$He (right panel) measured by the ALICE Collaboration in pp collisions at $\sqrt{s}= 7$ TeV~\cite{ALICE:2017xrp}. The blue curves are descriptions from the two-dimensional coalescence model described in the text.}
\label{d_he}
\end{figure}

Fig.~\ref{d_he} shows coalescence model results for the d and $^3$He in pp collisions at $\sqrt{s}=7$ TeV. 
In the model calculation, we also accounted for the correction arising from the relatively smaller system size in pp collisions compared to that of d and $^3$He\cite{Sun:2018mqq}. The correction factor can be derived from experimental data on the ratios of d/p and $^3$He/p between different collision systems~\cite{ALICE:2017xrp, ALICE:2015wav}.
The correction factor for deuteron to proton in pp collisions at $\sqrt{s}=7$ TeV ($\left< \mathrm{d}N_{ch}/ \mathrm{d}\eta \right>=6.01$) is 0.4065\cite{ALICE:2017xrp}, while that for  $^3$He/p at similar multiplicity is 0.13 \cite{ALICE:2021mfm}. 
These correction factors are multiplied after the model calculation using the following method. 

We used measured proton transverse momentum distribution in pp collisions at $\sqrt{s}=7$ TeV~\cite{ALICE:2015ial} multiplied by $r_f$ to remove the feed-down contribution in the formation of the deuteron and $^3$He.  For evaluating the feed-down fraction, we employed the statistical hadronization model (as expressed in Eq.~\ref{shm}) at a chemical freeze-out temperature of $181$ MeV. This temperature aligns with the freeze-out temperature deduced from the Blast-wave fit for pp collisions at 7 TeV and is anticipated to be analogous at 5.02 TeV. Then, one finds that $r_f=$26.6$\%$ of protons participate in the coalescence. The outcome achieved at $T=181$ MeV lies within the range of uncertainties stemming from our range of diquark masses 

Furthermore, we employ $A_{\rm{pp}}^{7 \ \rm{TeV}}=2.52$ fm$^2$. This value is derived by scaling the coalescence area in PbPb collisions at $\sqrt{s_{\rm{NN}}}=2.76$ TeV ($A_{\rm{PbPb}}^{2.76 \ \rm{TeV}}$=608 fm$^2$~\cite{Yun:2022evm}) with the charged particle multiplicity ratio between pp collisions at $\sqrt{s}=7$ TeV and PbPb collisions at $\sqrt{s_{\rm{NN}}}=2.76$ TeV.  
The charged particle multiplicity is $\left<\mathrm{d}N_{ch}/\mathrm{d}\eta \right>=6.01$ in pp collision at $\sqrt{s}=7$ TeV \cite{ALICE:2010mty} and $\left<\mathrm{d}N_{ch}/\mathrm{d}\eta \right>=1447.5 \pm 39$  in PbPb collisions at $\sqrt{s_{\rm{NN}}}=2.76$ TeV 0-10$\%$ event~\cite{ALICE:2013mez}. 

The consistency of our approach can be justified by the model description shown in Fig. \ref{d_he}, where we could have determined the two independent parameters $r_f$ and $A$ by fitting the experimental data for the deuteron and $^3$He, which turns out to yield almost the same values.  Hence, we will use $A=2.52$ fm$^2$ in Eq.~\eqref{xi_coal}.  

It is important to observe that both $V_H$ and $A$ scale in relation to the multiplicity, which is empirically proportional to $s^{0.103}$\cite{ALICE:2022wpn}. However, these quantities are used in Eq.~\eqref{xi_coal} as the ratio $V_H/A$. Consequently, even though we established this ratio at 7 TeV, its value is anticipated to be comparably consistent across pp collisions at both 5.02 and 13 TeV.

\subsection{Charm quark transverse momentum distribution}

\begin{figure}[h]
\centering
\includegraphics[width=.45\linewidth]{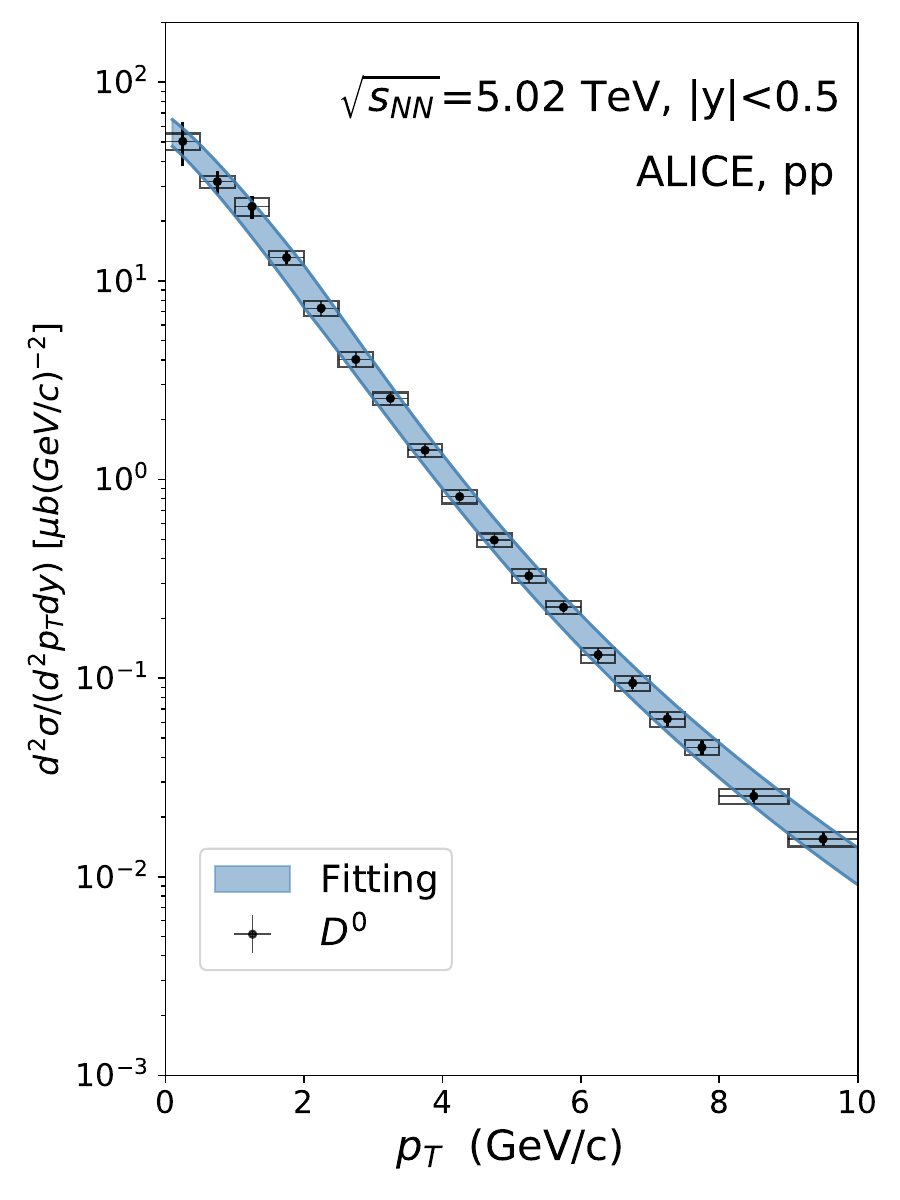}
\includegraphics[width=.45\linewidth]{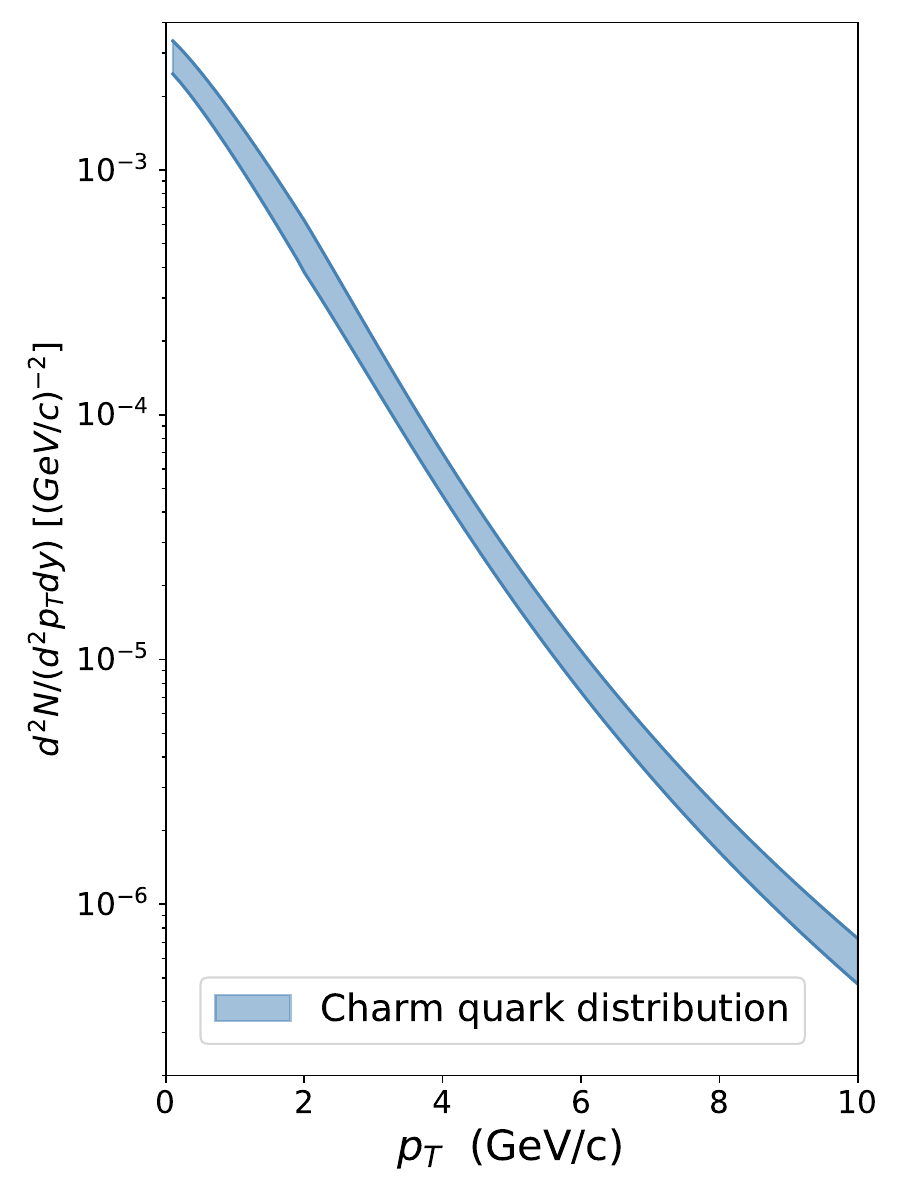}
\caption{Fitted $p_T$ distribution of $D^0$ and charm quark.  }
\label{charm_quark_distribution}
\end{figure}

To obtain the transverse momentum distribution of charm quark in pp collisions at $\sqrt{s}=5.02$ TeV, we used the transverse momentum distribution of $D^0$ meson \cite{ALICE:2019nxm} measured at the same collision energy. First, we assume that the shape of the charm quark distribution is the same as that of the $D^0$ meson. We then normalized this distribution using the ratio of the charm quark cross section $d\sigma^{c\bar{c}}/dy_{|y|<0.5}=1165 
$ $\mu$b \cite{ALICE:2021dhb} to fitted $D^0$ meson cross-section $d\sigma^{D^0}/dy_{|y|<0.5} =447$ $\mu$b~\cite{ALICE:2019nxm} .  Figure~(\ref{charm_quark_distribution}) shows the measured $D^0$ meson spectrum, the corresponding fit (left) and the charm quark distribution (right).

\subsection{Diquark distribution}

For the $(us)$ diquark $p_T$ distribution we adopt the blast-wave distribution. The shape of the diquark distribution was obtained from the blast-wave model given in Eq.~\eqref{equation:blastwave} with the normalization constant $\alpha$ to fit the statistical model prediction for the total number after integrating over the transverse momentum. Finally, the $p_T$ distribution of $\Xi_c^0$ composed of the $c$ quark and $(us)$ diquark becomes
\begin{widetext}
\begin{align}
\frac{\mathrm{d}^2 N_{\Xi_c^+}}{\mathrm{d}^2 P_{\mathrm{T}} \mathrm{d}y} =  \frac{g_{\Xi_c^+}}{g_{[us]} g_c}  \left( 2\sqrt{\pi} \sigma \right)^2 \frac{ \alpha }{2\pi A } \ \int_0^R  r\, {\rm d}r  \int d^2 p_{1\mathrm{T}} \ m_{1\mathrm{T}}
\times I_0\left( \frac{p_{1\mathrm{T}} \sinh \rho}{T_{kin}} \right) K_1 \left( \frac{m_{1\mathrm{T}} \cosh \rho}{T_{kin}} \right)  \int d^2 p_{2\mathrm{T}}\frac{\mathrm{d}^2 N_c}{ \mathrm{d}^2 p_{c\mathrm{T}} \mathrm{d}y} 
\nonumber \\
\times\exp{\left[ -\sigma^2 p'^2 \right]}  \delta^{(2)} ( \vec{P}_{\mathrm{T}} - \vec{p}_{\mathrm{1T}} - \vec{p}_{c\mathrm{T}} ).
\label{coal1}
\end{align}
\end{widetext}
Here, we use $A$=2.52 fm$^2$ 
and $V_H=20$ fm$^3$ determined at $\sqrt{s}=7$ TeV. However, since both quantities scale as multiplicity, we use the same ratio $V_H/A$ to calculate the $p_{\rm{T}}$ distribution of $\Xi_c$ in pp collisions at $\sqrt{s}=5.02$ TeV. Furthermore, it is important to mention that the smaller size of $\Xi_c$ allows us to disregard the correction factor arising from the small coalescence size effect as it amounts to less than 2$\%$ correction.

\subsection{Result}

\subsubsection{feed-down}

Since we are assuming $(qs)$ diquarks with spin 0 in the quark-gluon plasma, the coalescence with charm quark with non-zero relative orbital angular momentum $L$ will produce excited states with higher total spin. 
The coalescence of higher orbitals is sequentially suppressed, as shown in Ref.~\cite{ExHIC:2017smd}. Therefore, we have considered the contribution only from states generated with $L=1$, which will produce $\Xi_c$ excited states with negative parity and total spin 1/2 and 3/2, both of which are three-star states.  These states will contribute to the production of the $\Xi_c$ states through the feed-down process via the $\Xi_c'$ intermediate state, which eventually decay to $\Xi_c$ electromagnetically.

The yield ratios for the production of these excited states to the ground state $\Xi_c$ can be calculated by either the statistical hadronization model or the coalescence model, which results in similar ratios\cite{ExHIC:2017smd}.  Here, we use the ratios given by the statistical model at several temperatures used in the result section as given in Table \ref{tab:feed}.  

\begin{table}[h]
\centering
\begin{tabular}{c|c|c|c}
\hline \hline
$T$ (MeV) & $N_{\Xi_c(2790)} / N_{\Xi_c}$ & $N_{\Xi_c(2815)} / N_{\Xi_c}$ & Total \\  \hline 
165  & 0.1707 & 0.2971 & 0.4678 \\ \hline
181  & 0.2024 & 0.3569 & 0.5593\\ \hline 
184  & 0.2078 & 0.3673 & 0.5751\\ \hline  \hline
\end{tabular}
\caption{Feed-down fraction for selected chemical freeze-out temperatures.}
\label{tab:feed}
\end{table}
These ratios are added to the total production of the $\Xi_c$ state.  
Since coalescence from higher orbitals as well as higher nodal modes will be further suppressed, we can neglect contributions from any other possible excited states at this stage.

\subsubsection{Total contribution}

\begin{figure}[h]
\centering
\includegraphics[width=1\linewidth]{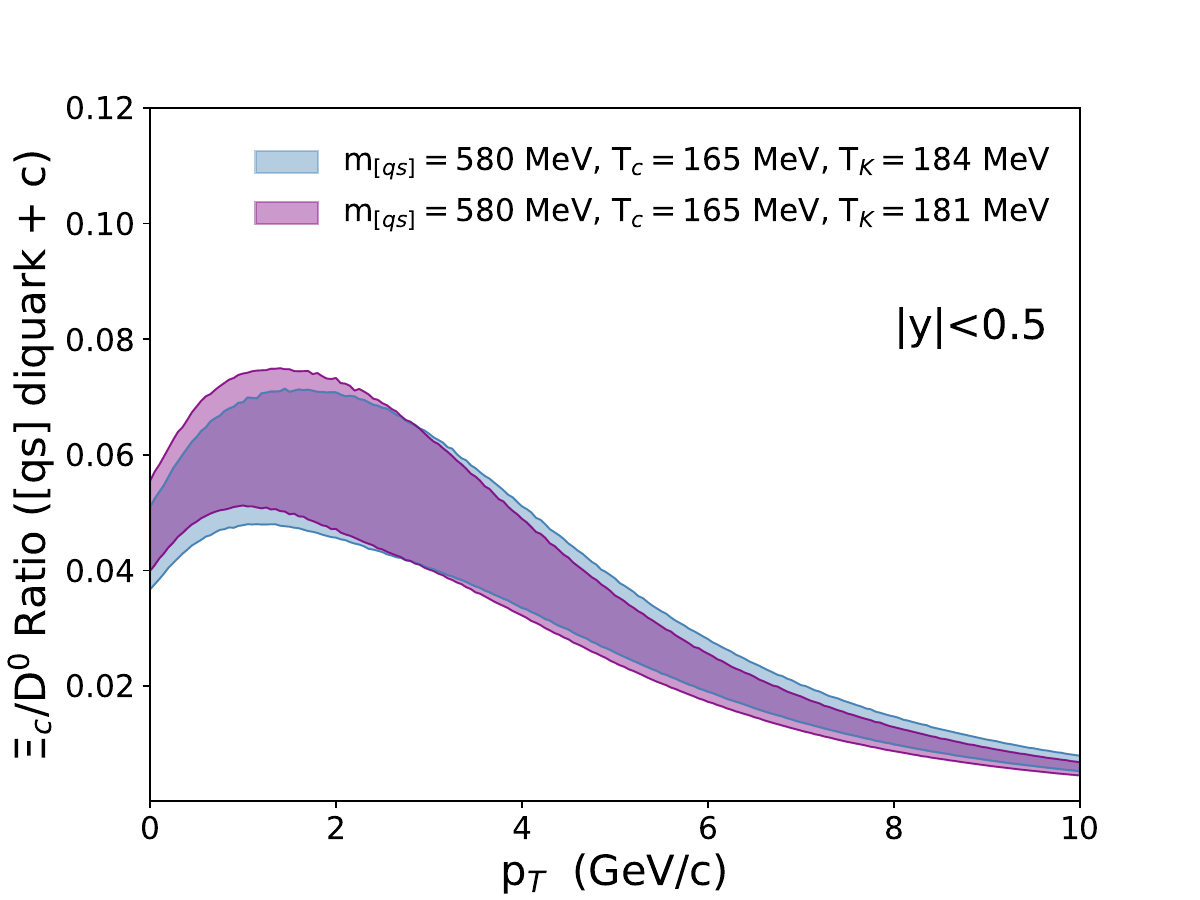}
\includegraphics[width=1\linewidth]{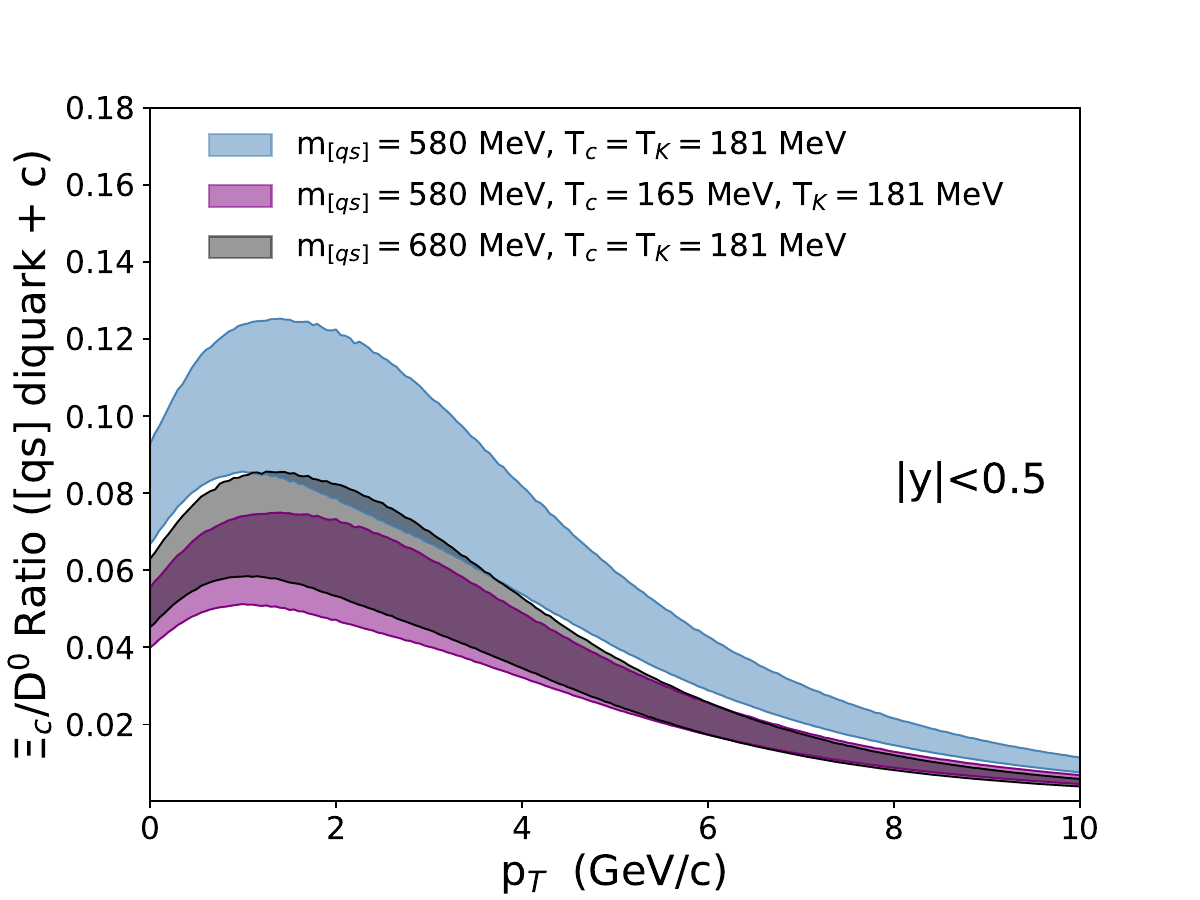}
\caption{$\Xi_c/D^0$ ratio due to $(qs)$ diquark with different masses . The upper panel shows the effects of different $T_K$, while the lower panel shows different combinations of $T_K$, $T_c$ and diquark mass. Although the ratio was obtained for 5.02 TeV, the value will only be slightly different for 13 or 7 TeV.}
\label{fig:diquark_ratio}
\end{figure}

In Figure~\ref{fig:diquark_ratio}, we present the transverse momentum distribution of the $\Xi_c^0$/$D^0$ ratio. The $p_\mathrm{T}$ distribution of $\Xi_c$ comprising a $(us)$ diquark and a charm quark is computed using Eq.~\eqref{coal1} for various $(us)$ diquark masses permitted by different schemes. Experimental data is employed for the $D^0$ meson, while the upper and lower bounds are derived from the associated experimental uncertainties in its distribution.  

To quantify the systematic uncertainty, we consider several scenarios for the chemical and kinetic freeze-out temperatures, denoted as $T_c \sim T_H$ and $T_K$, respectively. Additionally, we explore cases with different diquark masses. The temperature for kinetic decoupling is generally different from that for the phase transition where parton coalescence takes place because hadronic scattering delays the kinetic decoupling after hadronization.
But if the size of nuclear matter, or in other words, the particle multiplicity is small as in pp collisions, the effects of hadronic scattering will be less important. This is supported by the blast wave model, where the temperature for the kinetic freeze-out $T_K$ is not far from $T_c$.
Nevertheless, to probe the uncertainties associated with the effective kinetic freeze-out temperature, we compare the two possible kinetic freeze-out temperatures $T_K=184$ and 181 MeV as given in Table \ref{tab:values} for pp collision depending on collision energy.  As can be seen in the upper panel of Figure~\ref{fig:diquark_ratio}, the fit with a slightly larger temperature has a higher flow velocity, consequently shifting the distribution slightly towards higher $p_T$. In the lower figure, we 
consider three possible scenarios for the pp collision:
i) We first take $m_{[qs]}=580$ MeV and $T_c=T_K=181$ MeV as given by the blast-wave model.
ii) $T_C=165$ MeV and $T_K=181$ MeV as given in Ref.~\cite{Minissale:2020bif} and the blast-wave model, respectively.
iii) To probe the uncertainty in the diquark mass, we take $m_{[qs]}=680$ MeV and $T_c=T_K=181$ MeV. 
The lower Figure \ref{fig:diquark_ratio} shows the transverse momentum distribution for the $\Xi_c/D^0$ ratio calculated for the possible scenarios. As can be seen in the plot, the only noticeable difference comes from the values of $T_c$ and $m_{[qs]}$, which lead to variations in the overall magnitude of the production. One can observe the characteristic peak structure remains around $p_T \sim 2$ GeV.

\begin{figure}[h]
\centering
\includegraphics[width=1\linewidth]{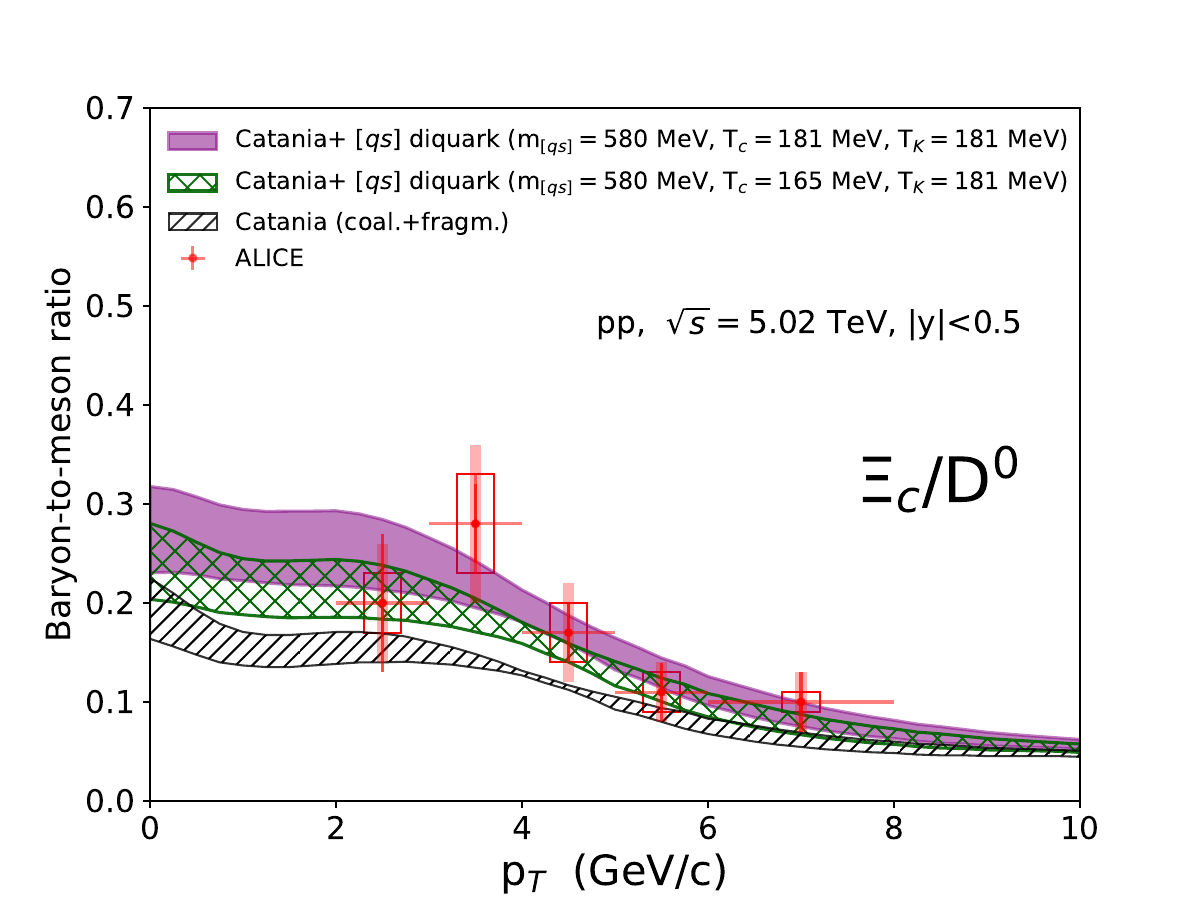}
\caption{$\Xi_c/D^0$ ratio after adding the enhancement due to the $(qs)$ diquark with mass 580 MeV to the coalescence and fragmentation calculation in Ref.~\cite{Minissale:2020bif}. }
\label{ratio_enhancement}
\end{figure}

Lastly, in Figure~\ref{ratio_enhancement}, we depict the outcome for a $(us)$ diquark mass of 580 MeV and $T_K=181$ MeV with two different $T_c=181$ and 165 MeV after incorporating the production ratios calculated in Ref. \cite{Minissale:2020bif}.
It should be noted that in Refs. \cite{Minissale:2020bif} and \cite{Cho:2019lxb}, the overall normalization in the coalescence formula for charmed hadrons is fixed by requiring that all zero transverse momentum charm quarks contribute to the production of the charmed hadrons. Therefore, if there is additional production of $\Xi_c$ originating from $(qs)$ diquarks, the normalization must be readjusted. However, charm coalescence is primarily driven by the production of $D$, $D^*$, $D_s$, $D_s^*$, and $\Lambda_c$ as discussed in Ref.~\cite{ExHIC:2017smd}. Consequently, we did not modify the normalization factor for the $\Xi_c/D^0$ ratio from Ref. \cite{Minissale:2020bif} when incorporating it into Figure 5.
Furthermore, while these calculations are derived for $\sqrt{s}=5.02$ TeV, any corrections anticipated in higher-energy pp collisions are likely to be negligible for the ratios.

As observed in the illustration, assuming a diquark mass of 580 MeV, the introduced contribution seems to provide the additional strength needed to reproduce the experimental observation.

\section{Discussion and summary}

We have demonstrated that within a quark model employing parameters that effectively reproduce ground state hadron masses, a $(us)$ diquark in the color anti-triplet spin-0 channel exhibits stronger binding compared with a $(ud)$ diquark in the same color and spin state.

This heightened attraction persists even when the potentials and constituent quark masses are adjusted to approximate values near the critical temperature. The presence of such robust $(us)$ or $(ds)$ diquarks is anticipated to amplify the production of charmed baryons where the dominant diquark configurations align with these "good" diquarks.

Furthermore, using the coalescence model, we have illustrated that such diquark correlations might potentially amplify $\Xi_c$ production, thereby offering a plausible explanation for the observed anomalous enhancement in $\Xi_c/D^0$ production ratios within high-energy pp collisions. This effect is postulated to arise from the existence of a $(us)$ diquark with an approximate mass of 580 MeV at the point of hadronization.

Analogous enhancements are predicted in pPb or PbPb collisions. Comprehensive estimations will be detailed in an upcoming publication.

\section*{Acknowledgements}
This work was supported by Samsung Science and Technology
Foundation under Project Number SSTF-BA1901-04, and by the Korea National Research Foundation under grant number No. 2023R1A2C3003023.
B.D. acknowledges the support from Bundesministerium f\"{u}r Bildung und Forschung through ErUM-FSP T01 (F\"{o}rderkennzeichen 05P21RFCA1). The work of A.P. was supported by the Korea National Research Foundation under the grant number 2021R1I1A1A01043019. S. Lim acknowledges support from the National Research Foundation of Korea grants funded by the Korean government under project number NRF-2008-00458.

\appendix

\section{Parametrization of the transverse momentum distribution}
An exponential function and a power-law type function were used for low $p_{\mathrm{T}}$ and high $p_{\mathrm{T}}$, respectively. Here, $p_1=1$ GeV and $|y|<0.5$.

\begin{align}
\frac{\mathrm{d}N}{\mathrm{d}^2p_{\mathrm{T}} \mathrm{d}y} &= a e^{-b(p_{\mathrm{T}}/p_1)^c } \qquad (p_{\mathrm{T}} < p_c), 
\nonumber \\
&= \frac{d}{\left[ 1 + (p_{\mathrm{T}}/p_0)^2 \right]^e } \qquad (p_{\mathrm{T}} > p_c)
\end{align}

\begin{widetext}

\begin{table}[h]
\centering
\begin{tabular}{ccccccccc}
\hline\hline
 Particle & $\sqrt{s}$ (GeV) & a (GeV)$^{-2}$  & b & c & d (GeV)$^{-2}$ & e & $p_0$ (GeV) & $p_c$ (GeV)  \\
\hline 
p$_l$ & 7 & $ 6.733 \times 10^{-2}$ & 1.904 & 1.456 & $ 3.477 \times 10^{-2} $ & 3.421 & 1.436 & 1.2 \\ 
p$_u$ & 7 & $ 7.738 \times 10^{-2}$ & 1.829 & 1.404 & $ 4.546 \times 10^{-2} $ & 3.462 & 1.438 & 1.2  \\  
$D^0_l$ & 5.02 & $1.004 \times 10^{-3}$ & 0.8601 & 1.156 & 4.851$\times 10^{-4}$ & 3.067 & 2.864 & 2.0  \\ 
$D^0_u$ & 5.02 & $1.360 \times 10^{-3}$ & 0.7729 & 1.172 & 8.940 $\times 10^{-4}$ & 2.967 & 2.644 & 2.0  \\
\hline \hline
\end{tabular}
\caption{The fitting parameters of p and $D^0$ in pp collisions. The subscripts in the first column correspond to the fits for the upper bound ($u$) and lower bound ($l$).}
\end{table}

\end{widetext}

\bibliographystyle{apsrev4-1}
\bibliography{refs}

\end{document}